\documentclass[12pt]{article}
\usepackage{amsmath}  
\input amssym

\begin{document}

\def\prg#1{\medskip{\bf #1}}     \def\ra{\rightarrow}
\def\lra{\leftrightarrow}        \def\Ra{\Rightarrow}
\def\nin{\noindent}              \def\pd{\partial}
\def\dis{\displaystyle}          \def\dfrac{\dis\frac}
\def\grl{{GR$_\Lambda$}}         \def\vsm{\vspace{-9pt}}
\def\Lra{{\Leftrightarrow}}      \def\ads3{AdS$_3$}
\def\ads3{{\rm AdS$_3$}}
\def\Leff{\hbox{$\mit\L_{\hspace{.6pt}\rm eff}\,$}}
\def\bull{\raise.25ex\hbox{\vrule height.8ex width.8ex}}
\def\Tr{\hbox{\rm Tr\hspace{1pt}}}
\def\as{{\rm as}}     \def\inn{\,\rfloor\,}
\def\ric{{(Ric)}}     \def\tmgl{\hbox{TMG$_\Lambda$}}
\def\nmg{{NMG}}       \def\nmgl{\hbox{NMG$_\Lambda$}}
\def\mb#1{\mbox{\boldmath{$#1$}}}
\def\hd{\hspace{2pt}{}^\star\hspace{-1pt}}

\def\cL{{\cal L}}     \def\cM{{\cal M }}    \def\cE{{\cal E}}
\def\cH{{\cal H}}     \def\hcH{\hat{\cH}}   \def\bcH{{\bar\cH}}
\def\cK{{\cal K}}     \def\hcK{\hat{\cK}}   \def\bcK{{\bar\cK}}
\def\cO{{\cal O}}     \def\hcO{\hat{\cO}}   \def\tR{{\tilde R}}
\def\cB{{\cal B}}     \def\bV{{\bar V}}     \def\heps{\hat\epsilon}
\def\cT{{\cal T}}     \def\D{{\Delta}}      \def\L{{\mit\Lambda}}
\def\hcO{{\hat\cO}}   \def\hcH{\hat\cH}     \def\bL{{\L_0}}
\def\bu{{\bar u}}     \def\bv{{\bar v}}     \def\bw{{\bar w}}
\def\cR{{\cal R}}     \def\bcR{{\bar\cR}}   \def\bz{{\bar z}}
\def\hcR{{\hat\cR}}

\def\G{\Gamma}        \def\S{\Sigma}
\def\a{\alpha}        \def\b{\beta}         \def\g{\gamma}
\def\d{\delta}        \def\m{\mu}           \def\n{\nu}
\def\th{\theta}       \def\k{\kappa}        \def\l{\lambda}
\def\vphi{\varphi}    \def\ve{\varepsilon}  \def\p{\pi}
\def\r{\rho}          \def\Om{\Omega}       \def\om{\omega}
\def\s{\sigma}        \def\t{\tau}          \def\eps{\epsilon}
\def\tom{{\tilde\om}} \def\nab{\nabla}
\def\tcR{\tilde{\cal R}} \def\tH{\tilde H}  \def\tG{\tilde G}
\def\Th{\Theta}       \def\cT{{\cal T}}     \def\cS{{\cal S}}
\def\cV{{\cal V}}     \def\hR{{\hat R}{}}
\def\f{{f}}           \def\ups{{\varphi}}
\def\nn{\nonumber}
\def\be{\begin{equation}}             \def\ee{\end{equation}}
\def\ba#1{\begin{array}{#1}}          \def\ea{\end{array}}
\def\bea{\begin{eqnarray} }           \def\eea{\end{eqnarray} }
\def\beann{\begin{eqnarray*} }        \def\eeann{\end{eqnarray*} }
\def\beal{\begin{eqalign}}            \def\eeal{\end{eqalign}}
\def\lab#1{\label{eq:#1}}             \def\eq#1{(\ref{eq:#1})}
\def\bsubeq{\begin{subequations}}     \def\esubeq{\end{subequations}}
\def\bitem{\begin{itemize}}           \def\eitem{\end{itemize}}
\renewcommand{\theequation}{\thesection.\arabic{equation}}

\title{Hamiltonian analysis of BHT massive gravity}

\author{M. Blagojevi\'c and B. Cvetkovi\'c\footnote{
        Email addresses: {\tt mb@ipb.ac.rs,
                                cbranislav@ipb.sc.rs}} \\
University of Belgrade, Institute of Physics,\\
P. O. Box 57, 11001 Belgrade, Serbia}
\date{}
\maketitle

\begin{abstract}
We study the Hamiltonian structure of the Bergshoeff-Hohm-Townsend
(BHT) massive gravity with a cosmological constant. In the space of
coupling constants $(\bL,m^2)$, our canonical analysis reveals the
special role of the condition $\bL/m^2\ne-1$. In this sector, the
dimension of the physical phase space is found to be $N^*=4$, which
corresponds to two Lagrangian degree of freedom. When applied to the
AdS asymptotic region, the canonical approach yields the conserved
charges of the BTZ black hole, and central charges of the asymptotic
symmetry algebra.
\end{abstract}

\section{Introduction}
\setcounter{equation}{0}

The new theory of massive gravity in three dimensions (3D), recently
proposed by Bergshoeff, Hohm and Townsend (BHT) \cite{1,2}, is defined
by adding the parity invariant, curvature-squared terms to the
Einstein-Hilbert action. With the cosmological constant $\bL$ and
the sign of the Einstein-Hilbert term $\s=\pm 1$, the action takes the
form
\be
I=a\int d^3 x\sqrt{g}\left(\s R-2\bL+\frac{1}{m^2}K\right)\, ,
\qquad K:=\hR_{ij}\hR^{ij}-\frac{3}{8}R^2\, ,              \lab{1.1}
\ee
where $\hR_{ij}$ is the Ricci tensor and $R$ the scalar curvature. At
the linearized level in asymptotically Minkowskian spacetime, the BHT
gravity is equivalent to the Pauli-Fierz theory for a free massive
spin-2 field. The action \eq{1.1} ensures the absence of ghosts
(negative energy modes), and the unitarity in flat space \cite{3};
moreover, the theory is renormalizable \cite{4}. In the AdS background
and for generic values of the coupling constants, the unitarity of the
massive gravitons is found to be in conflict with the positivity of
central charges in the boundary CFT \cite{5,2}. One should also note
that the BHT theory possesses a number of exact solutions \cite{6,7,8},
its AdS sector is studied in \cite{5,9}, central charges are discussed
in \cite{5,2,10}, and supersymmetric extension in \cite{11} .

It is interesting to observe that the particle content of the BHT
gravity depends on the values of coupling constants. Thus, if we
consider a maximally symmetric vacuum state defined by
$G_{ij}=\Leff\eta_{ij}$, where $G_{ij}$ is the Einstein tensor and
$\Leff$ the effective cosmological constant, this configuration
solves the BHT field equations if $\Leff$ solves a simple
quadratic equation. For $\bL/m^2=-1$, two solutions for $\Leff$
coincide, and the two massive modes degenerate with each other
\cite{2,5}. In that case, there is an extra \emph{gauge symmetry}
at the linearized level which allows massive modes to become
\emph{partially massless} \cite{2,12,x1}. The modes corresponding
to $\bL/m^2=3$ are also found to be special, but they remain
massive \cite{2}.

Motivated by the fact that the nature of physical modes in the BHT
gravity has been studied only in the \emph{linear approximation}, see
also \cite{13}, we use here the constrained Hamiltonian approach to
clarify the dynamical content of the BHT gravity
\emph{nonperturbatively}. In particular, we will find out a natural
role of the condition $\bL/m^2\ne-1$ in the canonical consistency
procedure.

The paper is organized as follows. In section 2, we give a brief
account of the basic dynamical features of the BHT gravity in the
Lagrangian formalism, and describe the BTZ black hole solution. In
section 3, we apply Dirac's method for constrained dynamical systems
\cite{14} to make a consistent canonical analysis of the BHT gravity.
Then, in section 4, we classify the constraints and find that the
theory exhibits two local Lagrangian degrees of freedom. To obtain this
result, we used a condition which, when applied to maximally symmetric
solutions, takes the form $\bL/m^2\ne -1$, corresponding to the case
of massive gravitons. In section 5, we find the form of the gauge
generator, showing thereby that the obtained classification of
constraints is correct. In section 6, we briefly describe the AdS
asymptotic structure by imposing the Brown-Henneaux asymptotic
conditions, find the form of the improved generators and the
corresponding conserved quantities, and calculate the central charges
of the asymptotic symmetry. Finally, section 7 is devoted to concluding
remarks, while appendices contain some technical details.

Our conventions are given by the following rules: the Latin indices
refer to the local Lorentz frame, the Greek indices refer to the
coordinate frame;  the middle alphabet letters
$(i,j,k,...;\m,\n,\l,...)$ run over 0,1,2, the first letters of the
Greek alphabet $(\a,\b,\g,...)$ run over 1,2; the metric components in
the local Lorentz frame are $\eta_{ij}=(+,-,-)$; totally antisymmetric
tensor $\ve^{ijk}$ and the related tensor density $\ve^{\m\n\r}$ are
both normalized as $\ve^{012}=1$.

\section{Lagrangian dynamics in the first order formalism}
\setcounter{equation}{0}

The BHT massive gravity with a cosmological constant is formulated as a
gravitational theory in Riemannian spacetime. Instead of using the
standard Riemannian formalism, with an action defined in terms of the
metric as in \eq{1.1}, we find it more convenient to use the triad
field and the spin connection as fundamental dynamical variables. Such
an approach can be naturally described in the framework of Poincar\'e
gauge theory \cite{15}, where basic gravitational variables are the
triad field $b^i$ and the Lorentz connection $A^{ij}=-A^{ji}$
(1-forms), and the corresponding field strengths are the torsion $T^i$
and the curvature $R^{ij}$ (2-forms). After introducing the notation
$A^{ij}=:-\ve^{ij}{_k}\om^k$ and $R^{ij}=:-\ve^{ij}{_k}R^k$, we have:
$$
T^i=db^i+\ve^i{}_{jk}\om^j\wedge b^k\, ,\qquad
R^i=d\om^i+\frac{1}{2}\,\ve^i{}_{jk}\om^j\wedge\om^k\, .
$$
The antisymmetry of $A^{ij}$ ensures that the underlying geometric
structure corresponds to Riemann-Cartan geometry, in which $b^i$ is
an orthonormal coframe, $g:=\eta_{ij}b^i\otimes b^j$ is the metric of
spacetime, $\om^i$ is the Cartan connection, and $T^i,R^i$ are the
torsion and the Cartan curvature, respectively. For $T_i=0$, this
geometry reduces to Riemannian. In what follows, we will omit the
wedge product sign $\wedge$ for simplicity.

The description of the BHT massive gravity can be technically
simplified as follows.
\bitem
\item[(a)] We use the triad field $b^i$ and the spin
connection $\om^i$ as independent dynamical variables.\vsm
\item[(b)] The Riemannian nature of the connection is ensured by imposing
the vanishing of torsion with the help of the Lagrange multiplier
$\l^i=\l^i{_\m}dx^\m$.\vsm
\item[(c)] Finally, by introducing an auxiliary field
$\f^i=\f^i{_\m}dx^\m$, we transform the term $K$ into an expression
linear in curvature.
\eitem
These modifications lead to a new formulation of the BHT massive
gravity, classically equivalent to \eq{1.1}:
\bsubeq\lab{2.1}
\be
L=a\left(2\s b^iR_i-\frac{1}{3}\bL\ve_{ijk}b^ib^jb^k
         +\frac{1}{m^2}L_K\right)+\l^iT_i\, .              \lab{2.1a}
\ee
Here, the piece $L_K$ is linear in curvature and depends on the
auxiliary field $\f^i$:
\be
L_K=R_i\f^i-V_K\, ,\qquad
V_K:=\frac{1}{4}\f_i\hd\left(\f^i-\f\,b^i\right)
    =\cV_K\,\heps\, ,                                      \lab{2.1b}
\ee
\esubeq
where $\f=\f^k{_k}$ and $\heps=b^0b^1b^2$ is the volume 3-form. In the
component notation, with $R_{imn}=G^k{_i}\ve_{kmn}$, $L_K$ takes the
well-known form \cite{1}:
$$
L_K=\left(\f_{ik}G^{ik}-\cV_K\right)\heps\,,
\qquad \cV_K:=\frac{1}{4}(\f_{ik}\f^{ik}-\f^2)\,.
$$
The form of $V_K=V_K(b^i,\f^i)$ ensures that after using the field
equations to eliminate $\f^i$, $L_K$ reduces to $K\heps$ (Appendix A).

\subsection*{The field equations}

Variation with respect to $b^i,\om^i,\f^i,\l^i,$ yields the BHT field
equations:
\bsubeq\lab{2.2}
\bea
&&a\left(2\s R_i-\bL\ve_{ijk}b^jb^k-\frac{1}{m^2}\Th_i\right)
  +\nabla\l_i=0\, ,                                        \\
&&a\left(2\s T_i+\frac{1}{m^2}\nabla\f_i\right)
  +\ve_{imn}\l^mb^n=0\, ,                                  \\
&&2R_i-\hd\left(\f^i-\f\,b^i\right)=0\, ,                  \\
&&T_i=0\,,
\eea
\esubeq
where $\Th_i=-\pd L_K/\pd b^i$ is the energy-momentum current (2-form)
associated to $L_K$, and $\nabla$ is the covariant derivative: for a
1-form $X_i$, $\nabla X_i=dX_i+\ve_{ijk}\om^j X^k$.

The last equation ensures that spacetime is Riemannian. The third
equation implies:
\bea
&&\f^i-\f b^i=\hd R^i=2G^i{_k}b^k\, ,                      \nn\\
&&2\f=R\, , \quad\f^i=2L^i=2L^i{_k}b^k\, ,                 \lab{2.3}
\eea
where $G_{ij}$ is the Einstein tensor, and $L_{ij}$ the Schouten
tensor:
\be
G_{ij}:=\hR_{ij}-\frac{1}{2}\eta_{ij}R\,,\qquad
L_{ij}:=\hR_{ij}-\frac{1}{4}\eta_{ij}R\, .                 \nn
\ee
Introducing the Cotton 2-form $C_i=\nabla L_i$, the second equation
reads
$$
\frac{2a}{m^2}C_i+\ve_{imn}\l^m b^n=0\, .
$$
Next, we introduce the Cotton tensor $C_{ij}$ by $C_i=C^k{_i}\heps_k$,
where $\heps_k=\frac{1}{2}\ve_{kmn}b^mb^n$, and note that the previous
equation, combined with $C^i{_i}=0$, implies:
\bea
&&\l_{ij}=\frac{2a}{m^2}C_{ij}\,,\qquad
  C_{ij}=\ve_i{}^{mn}\nabla_m L_{nj}\, ,                   \nn\\
&&\nabla\l_i=\frac{2a}{m^2}\left(\nabla_m C_{in}\right)b^mb^n\,.\nn
\eea
Now, the first field equation takes the form:
\bsubeq
\be
2\s R_i-\bL\ve_{imn}b^mb^n -\frac{1}{m^2}\Th_i
  +\frac{2}{m^2}\left(\nabla_m C_{in}\right)b^mb^n=0\, .  \lab{2.4a}
\ee
We can express the energy-momentum current $\Th_i$ in terms of the
corresponding energy-momentum tensor $\cT_i{^n}$ as (Appendix A)
$$
\Th_i=\cT_i{^n}\heps_n\, ,\qquad
\cT_i{^n}:=\d^n_i\cV_K-\frac{1}{2}\f_{ik}(\f^{kn}-\f\eta^{kn})\, .
$$
Expanding \eq{2.4a} in the dual basis $\heps^j$, with
$R_i=2G_{ij}\heps^j$, yields:
\be
\s G_{ij}-\bL\eta_{ij}-\frac{1}{2m^2}K_{ij}=0\, ,          \lab{2.4b}
\ee
where
\bea
K_{ij}&:=&\cT_{ij}-2(\nabla_m C_{in})\ve^{mn}{_j}          \nn\\
  &=&K\eta_{ij}-2L_{ik}G^k{_j}-2(\nabla_m C_{in})\ve^{mn}{_j}\,.\nn
\eea
\esubeq
These equations coincide with those found in \cite{5,2} (Appendix A).

We display here a set of algebraic consequences of the field equations:
\bsubeq
\bea
&&\f_{ij}=\f_{ji}\, ,                                      \lab{2.5a}\\
&&\l_{ij}=\l_{ji}\, ,\qquad \l=0\, ,                       \lab{2.5b}\\
&&\s\f+3\bL+\frac{1}{2m^2}\cV_K=0\, ,                      \lab{2.5c}
\eea
\esubeq
where we used $\cT_n{^n}=\cV_K$. Consider now a maximally symmetric
solution, for which
\be
R_{ijk}=\Leff\ve_{ijk}\, ,\qquad \hR_{ij}=-2\Leff\eta_{ij}\, ,
\qquad R=-6\Leff\, .                                       \lab{2.6}
\ee
Equation \eq{2.5c} with $\f_{km}=2L_{km}$ implies that the effective
cosmological constant $\Leff$ satisfies the quadratic equation
$$
\L^2_{\rm eff}+4m^2\s\Leff-4m^2\bL=0\, ,
$$
which yields
\be
\Leff=-2m^2\left(\s\pm\sqrt{1+\bL/m^2}\right)\, .       \lab{2.7}
\ee

\subsection*{BTZ black hole solution}

In the AdS sector of the BHT gravity, with $\Leff=-1/\ell^2$, there
exists a maximally symmetric solution, locally isomorphic to the BTZ
black hole \cite{1,16,17}.

In the Schwartzschild-like coordinates $x^\m=(t,r,\vphi)$, the BTZ
black hole solution is defined in terms of the lapse and shift
functions, respectively:
$$
N^2=\left(-8Gm_0+\frac{r^2}{\ell^2}
    +\frac{16G^2J_0^2}{r^2}\right)\, ,\qquad
N_\vphi=\frac{4GJ_0}{r^2}\, ,
$$
where $m_0,J_0$ are the integration parameters and $\Leff=-1/\ell^2$.
The triad filed has the simple diagonal form
\bsubeq\lab{2.8}
\be
b^0=Ndt\, ,\qquad b^1=N^{-1}dr\, ,\qquad
b^2=r\left(d\vphi+N_\vphi dt\right)\, ,                    \lab{2.8a}
\ee
while the Riemannian connection reads:
\be
\tom^0=-Nd\vphi\, ,\qquad \tom^1=N^{-1}N_\vphi dr\, ,\qquad
\tom^2=-\frac{r}{\ell^2}dt-rN_\vphi d\vphi\, .             \lab{2.8b}
\ee
Then, using \eq{2.6} and $C_{ij}=0$, the field equations imply that the
Lagrange multiplier $\l^i$ vanishes, while the auxiliary field $\f^i$
is proportional to the triad field:
\be
\l^i=0\, , \qquad \f^i=\frac{1}{\ell^2}b^i\, .             \lab{2.8c}
\ee
\esubeq

\section{Hamiltonian and constraints}
\setcounter{equation}{0}

In local coordinates $x^\m$, the component form of the Lagrangian
density reads:
\bsubeq
\be
\cL=a\ve^{\m\n\r}\left(\s b^i{_\m}R_{i\n\r}
    -\frac{1}{3}\bL\ve^{ijk}b_{i\m}b_{j\n}b_{\k\r}\right)
    +\frac{a}{m^2}\cL_K+\frac12\ve^{\m\n\r}\l^i{_\m}T_{i\n\r}\, ,
\ee
where the term $\cL_K$ is conveniently represented in the first order
formalism as
\be
\cL_K=\frac{1}{2}\ve^{\m\n\r}\f^i{_\m}R_{i\n\r}-b\cV_K\, ,
\ee
\esubeq
where $b=\det(b^i{_\m})$.

\prg{Primary constraints.} From the definition of the canonical momenta
$(\pi_i{^\m},\Pi_i{^\m},p_i{^\m},P_i{^\m})$, conjugate to the basic
dynamical variables ($b^i{_\m}, \om^i{_\m},\l^i{_\m},\f^i{_\m})$,
respectively, we obtain the primary constraints:
\bea
&&\phi_i{^0}:=\pi_i{^0}\approx 0\, ,\qquad\,\,
  \phi_i{^\a}:=\pi_i{^\a}-\ve^{0\a\b}\l_{i\b}\approx 0\, , \nn\\
&&\Phi_i{^0}:=\Pi_i{^0}\approx 0\, ,\qquad
  \Phi_i{^\a}:=\Pi_i{^\a}-2a\ve^{0\a\b}\left(
    \s b_{i\b}+\frac{1}{2m^2}\f_{i\b}\right)\approx 0\,.   \nn\\
&&p_i{^\m}\approx 0\, ,\hspace{61pt} P_i{^\m}\approx 0\, . \lab{3.2}
\eea
The PB algebra of the primary constraints is displayed in Appendix B.

After noting that the term $b\cV_K$ is \emph{bilinear} in the variables
$b^i{_0}$ and $\f^i{_0}$, one can conveniently represent the canonical
Hamiltonian as
\be
\cH_c= b^i{}_0\cH_i+\om^i{}_0\cK_i+\f^i{_0}\cR_i+\l^i{_0}\cT_i
         +\frac{a}{m^2}b\cV_K+\pd_\a D^\a\, ,              \nn\\
\ee
where
\bea
&&\cH_i=-\ve^{0\a\b}\left(a\s R_{i\a\b}
        -a\bL\ve_{ijk}b^j{}_\a b^k{}_\b+\nabla_\a\l_{i\b}\right)\,,\nn\\
&&\cK_i=-\ve^{0\a\b}\left(a\s T_{i\a\b}+\frac{a}{m^2}\nab_\a\f_{i\b}
        +\ve_{ijk}b^j{}_\a \l^k{}_\b\right) \, ,           \nn\\
&&\cR_i=-\frac{a}{2m^2}\ve^{0\a\b}R_{i\a\b}\, ,            \nn\\
&&\cT_i=-\frac{1}{2}\ve^{0\a\b}T_{i\a\b}\,,                \nn\\
&& D^\a=\ve^{0\a\b}\left[ \om^i{}_0\left( 2a\s b_{i\b}
        +\frac{a}{m^2}\f_{i\b}\right)+b^i{}_0 \l_{i\b}\right]\,.\nn
\eea

\prg{Secondary constraints.} Going over to the total Hamiltonian,
\be
\cH_T=\cH_c +u^i{}_\m\phi_i{}^\m+v^i{}_\m\Phi_i{}^\m
            +w^i{}_\m p_i{^\m}+z^i{_\m}P_i{^\m}\, ,        \nn
\ee
where $(u^i{_\m},v^i{_\m},w^i{_\m},z^i{_\m})$ are canonical
multipliers, we find that the consistency conditions of the primary
constraints $\pi_i{}^0$, $\Pi_i{}^0$, $p_i{}^0$ and $P_i{^0}$ yield the
secondary constraints:
\bea
&&\hcH_i:=\cH_i+\frac{a}{m^2}b\cT_i{^0}\approx 0\, ,       \nn\\
&&\cK_i\approx 0\, ,                                       \nn\\
&&\hcR_i:=\cR_i+\frac{a}{2m^2}b(\f_i{^0}-\f h_i{^0})\approx 0\,,\nn\\
&&\cT_i\approx 0\, .                                       \lab{3.3}
\eea
They correspond to the $\m=0$ components of the field equations
\eq{2.2}. Using the relation
\be
\cV_K = b^i{_0}\cT_i{^0}
         +\f^i{_0}\frac{1}{2}(\f_i{^0}-\f h_i{^0})\, ,     \nn
\ee
the canonical Hamiltonian can be rewritten in the form
\be
\cH_c= b^i{}_0\hcH_i+\om^i{}_0\cK_i+\f^i{_0}\hcR_i
      +\l^i{_0}\cT_i+\pd_\a D^\a\, .                       \lab{3.4}
\ee

The consistency of the remaining primary constraints
$X_A:=(\phi_i{}^\a, \Phi_i{}^\a, p_i{}^\a, P_i{}^\a)$ leads to the
determination of the multipliers
$(u^i{}_\a,v^i{}_\a,w^i{}_\a,z^i{_\a})$ (Appendix B). However, we find
it more convenient to continue our analysis in the reduced phase space
formalism. Using the second class constraints $X_A$, we can eliminate
the momenta $(\pi_i{^a},\Pi_i{^\a},p_i{^\a},P_i{^\a})$ and construct
the reduced phase space $R_1$, in which the basic nontrivial Dirac
brackets (DB) take the following form (Appendix B):
\bea
&&\{b^i{_\a},\l^j{_\b}\}^*_1=\eta^{ij}\ve_{0\a\b}\d\, ,  \qquad
\{\om^i{_\a},\f^j{_\b}\}^*_1=\frac{m^2}{a}\eta^{ij}\ve_{0\a\b}\d \nn\\
&&\{\l^i{_\a},\f^j{_\b}\}^*_1=-2m^2\s\eta^{ij}\ve_{0\a\b}\d\,.\lab{3.5}
\eea
The remaining DBs are the same as the corresponding Poisson brackets.

In $R_1$, the total Hamiltonian takes the simpler form:
\be
\cH_T=\cH_c +u^i{}_0\phi_i{}^0+v^i{}_0\Phi_i{}^0
            +w^i{}_0 p_i{^0}+z^i{_0}P_i{^0}\, ,
\ee
$\cH_c$ is given by \eq{3.4}, and the consistency conditions \eq{3.3}
remain unchanged.

\prg{Tertiary constraints.} The consistency conditions of the secondary
constraints can be written in the form:
\bea
\{\hcH_i,H_T\}^*_1&\approx& \frac{a}{m^2}\nabla_\m(b\cT_i{^\m})
  -\frac{1}{2}\ve_{imn}b(\f^{m\m}-\f h^{m\m})\l^n{_\m}\, , \nn\\
\{\cK_i,H_T\}^*_1&\approx& 0\,,\nn\\
\{\cT_i,H_T\}^*_1&\approx& -\frac{1}{2}b\ve_{ijk}\f^{jk}\,,\nn\\
\{\hcR_i,H_T\}^*_1&\approx& \frac{a}{2m^2}\nab_\m
  \left[b(\f_i{^\m}-\f h_i{^\m})\right]\, ,                \lab{3.7}
\eea
where, on the right-hand side, we use the \emph{symbolic} notation
$\dot\phi:=\{\phi, H_T\}_1^*$. The result is obtained with the help of
the canonical algebra of constraints, displayed in Appendix C. By using
$\nabla_\m(bh_i{^\m})\approx 0$, the divergence of $b\cT_i{^\m}$ can be
represented in the form
$$
\nabla_\m(b\cT_i{^\m})\approx
   \frac{1}{4}bh_i{^\m}\nab_\m(\f_{mn}\f^{mn}-\f^2)
  -\frac{1}{2}\nab_\m\left[b(\f_{ji}\f^{j\m}-\f\f^\m{_i})\right]\,.
$$

The third relation in \eq{3.7} yields the following tertiary
constraints:
\bsubeq
\bea
&&\th_{0\b}:=\f_{0\b}-\f_{\b0}\approx 0\,,\\
&&\th_{\a\b}:=\f_{\a\b}-\f_{\b\a}\approx 0\,.
\eea
\esubeq
They represent Hamiltonian counterparts of the Lagrangian relations
\eq{2.5a}.

To find an \emph{explicit} form of the consistency conditions for
$\hcH_i$ and $\hcR_i$, we have to replace the time derivatives
$\dot\phi$ by their canonical expressions $\{\phi,H_T\}$. To do that,
we introduce the following change of variables in $\cH_T$:
\be
\pi_i{^0}{}':=\pi_i{^0}+\f_i{^k}P_k{^0}\, ,\qquad
z^i{_0}{}':=z^i{_0}-\f^i{_k}u^k{_0}\, ,
\ee
whereupon the $(\pi_i{^0},P_i{^0})$ piece of $\cH_T$ takes the form
\bea
u^i{_0}\pi_i{^0}+z^i{_0}P_i{^0}&=&u^i{_0}\pi_i{^0}{}'
  +z^i{_0}{}'P_i{^0} \, .                                  \nn
\eea
Besides, we introduce the generalized multipliers
\bea
&&U^i{_\m}=u^i{_\m}+\ve^{imn}\om_{m0}b_{n\m}\, ,           \nn\\
&&Z^i{_\m}=z^i{_\m}+\ve^{imn}\om_{m 0}\f_{n\m}\, ,         \nn
\eea
which correspond, on-shell, to $\nab_0 b^i{_\m}$ and $\nab_0
\f^i{_\m}$, respectively; moreover, we define
$$
Z'{}^i{_\m}=Z^i{_\m}-\f^i{_m}U^m{_\m}\, .
$$

The consistency condition of $\hcR_i$, multiplied first by $b^i{_0}$
and then by $b^i{_\b}$, yields:
\bsubeq\lab{3.10}
\bea
&&U^\n{_\n}(\f_0{^0}-\f)-\f_0{^\m}U^0{_\m}-(Z'{}^\a{_\a}-\f U^0{_0})
  +b^{-1}b^i{_0}\nabla_\a
      \left[b(\f_i{^\a}-\f h_i{^\a})\right]=0\, ,          \lab{3.10a}\\
&&g^{0\m}Z'_{\b\m}+\f_\b{^0}U^\a{_\a}
  -(\f_\b{^\a}-\f\d_\b^\a)U^0{_\a}
  +b^{-1}b^i{_\b}\nab_\a\left[b(\f_i{^\a}-\f
  h_i{^\a})\right]=0\, .                                   \lab{3.10b}
\eea
\esubeq
The first relations, in which the arbitrary multipliers $U_{k0}$ and
$Z'_{k0}$ are cancelled, contains only the determined multipliers
$U_{k\a}$ and $Z'_{k\a}$. Using the expressions for $U_{k\a}$ and
$Z'_{k\a}$ calculated with the help of Appendix B, one finds that this
relation reduces to an identity (Appendix D). The second relation
defines the two components $Z'_{\b 0}=b^k{_\b}Z'_{k 0}$ of $Z'_{k0}$.

The consistency condition of $\hcH_i$ in conjunction with \eq{3.10}
yields:
\bea
&&(\f^{j0}h_i{^\a}-\f^{j\a}h_i{^0})Z'_{j\a}
  +\f^{j\a}\nab_\a \f_{ji}-\f^{jk}h_i{^\a}\nab_\a\f_{jk}
  +\frac{m^2}{a}\ve_{ijk}(\f^{jn}-\f\eta^{jn})\l^k{_n}\approx 0\,.\nn
\eea
Substituting here the expression for the determined multiplier
$Z'_{j\a}$, we find:
$$
\f\ve_{ijk}\l^{jk}=0\, .
$$
Thus, for $\f\neq 0$, we obtain three tertiary constraints:
\bsubeq
\bea
&&\psi_{0\b}:=\l_{0\b}-\l_{\b0}\approx 0\,,\\
&&\psi_{\a\b}:=\l_{\a\b}-\l_{\b\a}\approx 0\,.
\eea
\esubeq
\bitem
\item[--] The consistency conditions of the secondary constraints
determine $Z'_{\b 0}$ and produce the tertiary constraints $\th_{\m\n}$
and $\psi_{\m\n}$.
\eitem

\prg{Quartic constraints.} The consistency of $\ve^{0\a\b}\th_{\a\b}$
reads
$$
\{\ve^{0\a\b}\th_{\a\b},H_T\}_1^*\approx \frac{4m^2}{a}b\l\approx 0\,,
$$
and we have a new, quartic constraint, the canonical counterpart of
\eq{2.5b}:
\bsubeq
\be
\chi:=\l\approx 0\, .
\ee
The consistency condition of $\th_{0\b}$ is identically satisfied
(Appendix D):
\be
\{\th_{0\b},H_T\}_1^*=z'_{0\b}-z'_{\b 0}\approx 0\, .      \lab{3.12b}
\ee
\esubeq
The consistency of $\ve^{0\a\b}\psi_{\a\b}$ reads:
$$
\{\ve^{0\a\b}\psi_{\a\b},H_T\}_1^*\approx
-4ab\left(\s\f+3\bL+\frac{1}{2m^2}\cV_K\right)\approx 0\,.
$$
Thus, we have a new quartic constraint:
\bsubeq
\be
\ups:=\s\f+3\bL+\frac{1}{2m^2}\cV_K\approx 0\,,
\ee
as expected from \eq{2.5c}.

To interpret the consistency condition for $\psi_{0\b}$, we introduce
the notation
\be
\pi_i{^0}{}'':=\pi_i'{^0}+\l^k{_i}p_k{^0}\, ,\qquad
w^i{_0}{}':=w^i{_0}-u^k{_0}\l_{ik}\, .                     \nn
\ee
The, the $(\pi_i{^0},P_i{^0},p_i{^0})$ piece of the Hamiltonian takes
the form
\bea
u^i{_0}\pi_i{^0}+w^i{_0}p_i{^0}+z^i{_0}P_i{^0}
 &=&u^i{_0}\pi_i{^0}{}''+w^i{_0}'p_i{^0}
    +z^i{_0}{}'P_i{^0} \, ,                                \nn
\eea
and we have:
\be
\{\psi_{0\b},H_T\}_1^*= w'_{0\b}-\bar w'_{\b 0}\approx 0\, .
\ee
\esubeq
Hence, the multipliers $w'_{\b0}$ are determined.
\bitem
\item[--] The consistency conditions of the tertiary constraints
determine $w'_{0\b}$ and produce the quartic constraints $\chi$ and
$\ups$.
\eitem

\prg{End of the consistency procedure.} The consistency condition
of the quartic constraint $\chi$ determines the multiplier $w'_{00}$:
\bea
&&\{\chi,H_T\}_1^*= w'^\m{_\m}\approx 0\,,\nn\\
&&g^{00}w'_{00}+g^{0\b}\bw'_{\b0}+h^{i\a}\bw'_{i\a}=0\,,
\eea
where $\bw'_{i\a}=\bw_{i\a}-\l_{ik}\bu^k{_\a}$.

The consistency condition for the quartic constraint $\ups$
has the form:
\bea
&&\{\ups,H_T\}_1^*=\Om^{\m\n}z'_{\m\n} \approx 0\, ,   \nn\\
&&\Om^{\m\n}:=\s g^{\m\n}
  +\frac{1}{4m^2}\left(\f^{\m\n}-\f g^{\m\n}\right)\, .
\eea
This relation determines the multiplier $z'_{00}$, provided the
coefficient $\Om^{00}$ does not vanish.
\bitem
\item[--] The consistency conditions for the quartic constraints
determine $w'_{00}$ and $z'_{00}$.
\eitem
This finally completes the consistency procedure. At the end, we wish
to stress that the completion of this process is achieved by employing
the following \emph{extra conditions}:
\bsubeq\lab{3.16}
\bea
&&\f\neq 0\, ,                                             \lab{3.16a}\\
&&\Om^{00}\neq 0\, .                                       \lab{3.16b}
\eea
\esubeq
Dynamical interpretation of these conditions is discussed in the next
section.

\section{Classification of constraints}
\setcounter{equation}{0}

Among the primary constraints, those that appear in $\cH_T$ with
arbitrary multipliers are first class (FC):
\bsubeq
\be
\pi_i{^0}{}'',\Pi_i{^0}=\mbox{FC}\, ,
\ee
while the remaining ones, $p_i{^0}$ and $P_i{^0}$, are second class.

Going to the secondary constraints, we use the following simple
theorem:
\bitem
\item[$\bull$] If $\phi$ is a FC constraint, then $\{\phi,H_T\}^*_1$
is also a FC constraint.
\eitem
The proof relies on using the Jacoby identity. The theorem implies that
the secondary constraints $\bcH_i:=-\{\pi_i{^0}{}'',H_T\}^*_1$ and
$\bcK_i=-\{\Pi_i{^0},H_T\}^*_1$ are FC. After a straightforward but
lengthy calculation, we obtain:
\bea
&&\bcH_i=\hcH''_i+h_i{^\m}(\nab_\m \l_{jk})b^k{_0}p^{j0}
  +h_i{^\m}(\nab_\m \f_{jk})b^k{_0}P^{j0}\, ,              \nn\\
&&\bcK_i=\cK_i-\ve_{ijk}(\l^j{_0}p^{k0}-b^j{_0}\l^k{_n}p^{n0})
  -\ve_{ijk}(\f^j{_0}P^{k0}-b^j{_0}\f^k{_n}P^{n0})\, ,
\eea
\esubeq
where $\hcH''_i:=\hcH_i+\f^k{_i}\hcR_k+\l^k{_i}\cT_k$. As before, the
time derivative $\dot\phi$ is a short for $\{\phi,H_T\}^*_1$.

The total Hamiltonian can be expressed in terms of the FC constraints
(up to an ignorable square of constraints) as follows:
\be
\hcH_T=b^i{_0}\bcH_i+\om^i{_0}\bcK_i
  +u^i{_0}\pi_i{^0}{}''+v^i{_0}\Pi_i{^0}\,.
\ee

In what follows, we will show that the complete classification of
constraints in the reduced space $R_1$ is given as in Table 1, provided
the conditions \eq{3.16} are satisfied.
\begin{center}
\doublerulesep 1.8pt
\begin{tabular}{lll}
\multicolumn{3}{l}{\hspace{16pt}Table 1. Classification
                                         of contraints in $R_1$} \\
                                                      \hline\hline
\rule{0pt}{12pt}
&~First class \phantom{x}&~Second class \phantom{x} \\
                                                      \hline
\rule[-1pt]{0pt}{15pt}
\phantom{x}Primary &~$\p_i{^0}{}'',\Pi_i{^0}$
            &~$p_i{^0},P_i{^0}$   \\
                                                      \hline
\rule[-1pt]{0pt}{15pt}
\phantom{x}Secondary\phantom{x} &~$\bcH_i,\bcK_i$
           &~$\cT_i,\hcR'_i$       \\
                                                      \hline
\rule[-1pt]{0pt}{15pt}
\phantom{x}Tertiary\phantom{x}
                  & &~$\th_{0\b},\th_{\a\b},\psi_{0\b},\psi_{\a\b}$ \\
                                                      \hline
\rule[-1pt]{0pt}{15pt}
\phantom{x}Quartic\phantom{x}
                 & &~$\chi,\ups$  \\
                                                      \hline\hline
\end{tabular}
\end{center}
Here, $\hcR'_i$ is a suitable modification of $\hcR_i$, defined so that
it does not contain $\f_{i0}$:
$$
\hcR'_i=\cR_i+\frac{ab}{2m^2}\left[(g^{00}h_i{^\a}
  -g^{0\a}h_i{^0})\f_{0\a}+g^{0\a}\f_{i\a}-h_i{^0}\f^\a{_\a}\right]\,.
$$

To prove the content of Table 1, we need to verify the second-class
nature of the constraints in the last column. This can be done by
calculating the determinant of their DBs. In order to simplify the
calculation, we divide the procedure into three simpler steps, as
described in Appendix E: (i) we start with the subset of 6 constraints
$Y_A:=(\th_{0\b},\ups,P^{\a 0}, P_0{^0})$ and show that they are
second class since the determinant of $\{Y_A,Y_B\}^*_1$ is nonsingular;
then, (ii) we extend our considerations to $Z_A:=(\psi_{0\b},\chi,
p^{\a 0}, p_0{^0})$, and show that these 6 constrains are also second
class; finally, (iii) we show in the same manner that the remaining 8
constraints $W_A:=(\cT_i,\hcR'_i,
\frac{1}{2}\ve^{0\a\b}\psi_{\a\b},\frac{1}{2}\ve^{0\a\b}\th_{\a\b})$
are second class.
\bitem
\item[$\bull$] Thus, all 20 constraints $(Y_A,Z_B,W_C)$ are second class.
\eitem
Note, however, that this result is valid only \emph{if the condition
\eq{3.16b} is satisfied}, as shown in Appendix E.

When the classifcation of constraints is complete, the number of
independent dynamical degrees of freedom in the phase space $R_1$ is
given by the formula:
$$
N^* = N-2N_1-N_2\, ,
$$
where $N$ is the number of phase space variables in $R_1$, $N_1$ is the
number of FC, and $N_2$ the number of second class constraints.
According to the results in Table 1, we have $N = 4\times 9+ 4\times
3=48$ ($4\times 6$ momentum variables are already eliminated from
$R_1$), $N_1 = 12$ and $N_2 = 20$. Consequently:
\bitem
\item[$\bull$] the number of physical modes in the phase space $R_1$ is
$N^* = 4$, and the theory exhibits 2 local Lagrangian degree of
freedom.
\eitem

What is the dynamical meaning of the extra conditions \eq{3.16} ? To
clarify this issue, let us consider their content for maximally
symmetric solutions.

When the first condition is violated, that is when $\f=0$, we
have $R=0$, $\Leff=0$, and $\hR_{ij}=0$. This is possible only when
$\bL=0$, as follows from the field equation \eq{2.4b}, and we have a
completely trivial dynamics. This motivates us to accept $\f\ne 0$ as a
natural dynamical assumption.

Turning to the second condition, we use $\f_{\m\n}=2 L_{\m\n}$ to
rewrite $\Om^{\m\n}$ in the form
$$
\Om^{\m\n}=\s g^{\m\n}+\frac{1}{2m^2}G^{\m\n}
 =g^{\m\n}\left(\s+\frac{1}{2m^2}\Leff\right)\, .
$$
Thus, $\Om^{00}$ vanishes when $\Leff=-2m^2\s$, or equivalently,
when $\bL/m^2=-1$, as follows from \eq{2.7}. At this point, the
mass spectrum of the BHT gravity, in the linearized approximation,
undergoes a serious transition, whereby the massive sector of
gravitons becomes partially \emph{massless}  \cite{2,12,x1}. At
the canonical level, this phenomenon is reflected in the fact
that, for $\Om^{00}=0$, the multiplier $z'_{00}$ remains
undetermined, and consequently, some of the second class
constraints become \emph{first class}. Thus, using $\Om^{00}\ne
0$, we stay in the massive sector of the BHT gravity. In
particular, the special case $\bL/m^2=3$ also belongs to this
sector. The canonical structure of the complementary sector
$\Om^{00}=0$ is left for future studies.

\section{Gauge generator}
\setcounter{equation}{0}

After completing the Hamiltonian analysis, we now employ the Castellani
procedure \cite{18} to construct the canonical gauge generator.
Starting with the primary FC constraints $\pi_i{^0}{}''$ and
$\Pi_i{^0}$, we find:
\bea
&&G[\t]=\dot\t^i\pi_i{^0}{}''
  +\t^i\left[\bcH^i-\ve_{ijk}\om^j{_0}\pi^k{^0}''
 -\ve_{ijk}b^j{_0}(\f^{kn}-\f\eta^{kn})\Pi_n{^0}\right]\, ,      \nn\\
&&G[\s]=\dot\s^i \Pi_i{^0}+\s^i\left(\bcK^i
  -\ve_{ijk}\om^j{_0}\Pi^k{_0}-\ve_{ijk}b^j{_0}\pi^{k0}{}''\right)\,.
\eea
The complete gauge generator has the form $G=G[\t]+G[\s]$, its action
on the fields is defined by the DB operation $\d_0\phi=\{\phi,G\}_1^*$,
but the resulting gauge transformations do not have the Poincar\'e
form. The standard Poincar\'e content of the gauge transformations is
obtained by introducing the new parameters \cite{19}
$$
\t^i=-\xi^\r b^i{_\r}\, ,\qquad \s^i=-\th^i-\xi^\r \om^i{_\r}\, .
$$
Expressed in terms of these parameters (and after neglecting some
trivial terms, quadratic in the constraints), the gauge generator
takes the form:
\bea
G&=&-G_1-G_2\, ,                                           \nn\\
G_1&=&\dot\xi^\m\left(b^i{}_\m\pi_i{}^0
       +\om^i{}_\m\Pi_i{}^0+\l^i{_\m}p_i{}^0
       +\f^i{_\m}P_i{^0}\right)                            \nn\\
&&+\xi^\m\left[b^i{}_\m\hcH_i+\om^i{}_\m\cK_i
               +\l^i{_\m}\cT_i+\f^i{_\m}\hcR_i\right.      \nn\\
&&\qquad+\left.(\pd_\m b^i{_0})\pi_i{}^0+(\pd_\m\om^i{}_0)\Pi^i{}_0
  +(\pd_\m\l^i{_0})p_i{^0}+(\pd_\m\f^i{_0})P_i{^0}\right]\,,\nn\\
G_2&=&\dot{\th^i}\Pi_i{}^0+\th^i\left[\cK_i
  -\ve_{ijk}\left(b^j{}_0\pi^{k0}+\om^j{}_0\Pi^{k0}
  +\l^j{}_0p^{k0}+\f^j{}_0P^{k0}\right)\right]\, .
\eea
Looking at the related gauge transformations, we find a complete
agreement with the Poincar\'e gauge transformations  \emph{on
shell}.

\section{Asymptotic structure in the AdS sector}
\setcounter{equation}{0}

Asymptotic conditions imposed on dynamical variables determine the form
of asymptotic symmetries, and consequently, they are closely related to
the gravitational conservation laws. In this section, we focus our
attention to the AdS sector of the theory, with $\Leff=-1/\ell^2$.

\prg{Asymptotic conditions.} The AdS asymptotic conditions are defined
by demanding that (a) the asymptotic configurations include the BTZ
black hole solution, (b) they are invariant under the action of the AdS
group $SO(2,2)$, and (c) the corresponding conserved charges are well
defined. These requirements are realized by the Brown-Henneaux type of
asymptotic conditions on the triad field $b^i{_\m}$ and the Riemannian
connection $\om^i{_\m}$, which have the same form as in the
topologically massive gravity \cite{19}. In the BHT massive gravity,
there are two more Lagrangian variables, the Lagrange multiplier
$\l^i{_\m}$ and the auxiliary field $\f^i{_\m}$, whose asymptotic
behavior is defined by generalizing \eq{2.8c}:
\be
\l^i{_\m}=\hcO\, ,\qquad
\f^i{_\m}=\frac{1}{\ell^2}b^i{_\m}+\hcO\,,               \lab{6.1}
\ee
where $\hcO$ denotes terms with arbitrarily fast asymptotic decrease.

Having chosen the asymptotic conditions, one should find the subset of
gauge transformations that respect these conditions. It turns out that
the parameters of the restricted gauge transformations are defined in
terms of two functions, $T^+(x^+)$ and $T^-(x^-)$, in the same way as
in \cite{19}. The resulting asymptotic symmetry of spacetime coincides
with the conformal symmetry.

\prg{The improved generator.} The canonical generator acts on dynamical
variables via the Dirac bracket operation, hence, it should have
well-defined functional derivatives. In order to ensure this property,
we have to improve the form  of $G$ by adding a suitable surface term
$\G$, such that $\tG=G+\G$ is a well-defined canonical generator. The
surface term of the improved canonical generator $\tG$ takes the form
\bsubeq
\be
\G=-\int_0^{2\pi}d\vphi\left(\xi^0\cE^1+\xi^2\cM^1\right)\,,
\ee
where
\bea
&&\cE^\a:=2a\ve^{0\a\b}\left[\left(\s+\frac{1}{2m^2\ell^2}\right)\om^0{_\b}
  +\frac 1{\ell}b^2{_\b}+\frac{1}{2m^2\ell}\f^2{_\b}\right]b^0{_0}\, ,             \nn\\
&&\cM^\a:=-2a\ve^{0\a\b}\left[\left(\s+\frac{1}{2m^2\ell^2}\right)\om^2{_\b}
  +\frac 1{\ell}b^0{_\b}+\frac{1}{2m^2\ell}\f^0{_\b}\right]b^2{_2}\, .
\eea
\esubeq

\prg{Conserved charges.} The values of the surface terms, calculated
for $\xi^0=1$ and $\xi^2=1$, define the energy and angular momentum of
the system, respectively. In particular, the energy and angular
momentum for the BTZ black hole are:
\be
E=\left(\s+\frac{1}{2m^2\ell^2}\right)m_0\, ,\qquad
M=\left(\s+\frac{1}{2m^2\ell^2}\right)J_0\, .
\ee
This results is verified by using Nester's general covariant formalism
\cite{20}, see also \cite{6}.

\prg{Central charges.} Using the notation
$\tG_{(i)}:=\tG[T_i^+,T_i^-]$, the main theorem of \cite{21} states
that the canonical algebra of the improved generators has the general
form:
\be
\left\{\tG_{(2)},\tG_{(1)}\right\}= \tG_{(3)}+C_{(3)}\, ,
\ee
where $C_{(3)}$ is the central term. Introducing the Fourier modes for
the improved generator, $L^\mp_n=-\tG[T^\mp=e^{inx^\mp}]$, the above
canonical algebra is found to have the form of two independent Virasoro
algebras with identical central charges,
\be
c^-=c^+=\frac{3\ell}{2G}\left(\s+\frac{1}{2m^2\ell^2}\right)\,,\lab{6.5}
\ee
see \cite{5,2,6}. Once we have the central charges, we can use Cardy's
formula to calculate the black hole entropy:
\be
S=\left(\s+\frac{1}{2m^2\ell^2}\right)\frac{2\pi r_+}{4G}\,,
\ee
where $r_+$ is the radius of the outer black hole horizon.

In order to have a unitary boundary CFT, the central charge \eq{6.5}
has to be positive. On the other hand, one also expects that massive
gravitons, defined as small excitations around the AdS background, should
carry positive energy.
Now, relying on the 
analysis performed in \cite{2}, one can conclude that for generic
values of the coupling constants, these two requirements are in
conflict with each other. For possible resolutions of this conflict,
see \cite{2}. Note, however, that the positivity of the central charge
and the BTZ black hole energy are in agreement with each other.

\section{Concluding remarks}

In this paper, we studied the BHT massive gravity as a constrained
dynamical system.

Our basic goal was to obtain and classify the constraints and deduce
the dimension of the physical phase space $N^*$. In the process of
completing Dirac's consistency procedure, we discovered the essential
role of the extra condition $\Om^{00}\ne 0$, Eq. \eq{3.16b}. When
applied to maximally symmetric solutions, this condition describes the
sector of massive gravitons with $\bL/m^2\ne -1$.
In this sector, the dimension of the phase space is found to be
$N^*=4$, which means that the theory exhibits 2 Lagrangian degrees of
freedom. The canonical structure of the complementary sector
$\Om^{00}=0$ with partially massless gravitons is left for future
studies.

As a particular application of our results, we examined the AdS
asymptotic structure of the theory. Using the Brown-Henneaux type of
asymptotic conditions, we calculated energy and angular momentum of the
BTZ black hole, and central charges of the asymptotic symmetry algebra.
Our results are in agreement with those existing in the literature.

\section*{Acknowledgements}

This work was supported by the Serbian Science Foundation under
Grant No. 141036. One of us (B.C.) would like to thank Daniel
Grumiller and his collaborators for useful discussions on
partially massless modes.

\appendix
\section{On the first order form of \boldmath{$L_K$}}
\setcounter{equation}{0}

In this appendix, we display several interesting relations related to
the first order formulation of $L_K$, defined in \eq{2.1b}.

The variation of $L_K$ with respect to $\f_i$ yields
$2R_i-\hd\left(\f^i-\f b_i\right)=0$. This equation can be solved for
$\f^i$ as in \eq{2.3}, which implies
\be
L_K=\frac{1}{2}R_i\f^i=R_iL^i=G_{ki}L^{ik}\heps=K\heps\, .
\ee
Thus, the expression for $L_K$ is classically equivalent to $K$.

Following the analogy with electrodynamics, we rewrite
the term $V_K$ in $L_K$ as:
\be
V_K=\frac{1}{2}\f_i\cH^i=\cV_K\heps\, ,\qquad
\cH^i:=\frac{1}{2}\hd\left(\f^i-\f b^i\right)\, .          \nn
\ee
The energy-momentum current (density) associated to $L_K$ is given by:
\be
\Th_i:=-\frac{\pd L_K}{\pd b^i}
  =\frac{\pd V_K}{\pd b^i}=b_i\inn V_K-\cH^k (b_i\inn\f_k)\, .
\ee
Then, using
$$
b_i\inn V_K=\cV_K\heps_i\, ,\qquad
  \cH^k(b_i\inn\f_k)
  =\frac{1}{2}\f_{ki}(\f^{kn}-\f\eta^{kn})\heps_n\, ,      \nn\\
$$
we find:
\be
\Th_i=\cT_i{^n}\heps_n\, ,\qquad
\cT_i{^n}:=\d^n_i\cV_K-\frac{1}{2}\f_{ki}(\f^{kn}-\f\eta^{kn})\,,
\ee
where $\cT_i{^n}$ is the dynamic energy-momentum tensor:
$$
\frac{\pd}{\pd b^i{_\m}}(b\cV_K)=b\cT_i{^\m}\, .
$$

Using the relations
\bea
&&\ve_j{}^{mn}\nabla_m C_{ni}
  =\nabla^m\nabla_jL_{mi}-\nabla^m\nabla_m L_{ji}
  =\hR_{ik}\hR^k{_j}-R_{imjn}\hR^{mn}
  +\frac{1}{4}\nabla_j\nabla_i R-\nabla^2L_{ij}\, ,        \nn\\
&&R_{imjn}\hR^{mn}=
  \eta_{ij}\left(\hR_{mn}\hR^{mn}-\frac{1}{2}R^2\right)
  +\frac{3}{2}\hR_{ij}R-2\hR_{in}\hR^n{_j}\, ,             \nn\\
&&L_{ik}G^k{_j}=\hR_{ik}\hR^k{_j}-\frac{3}{4}R\hR_{ij}
                +\frac{1}{8}\eta_{ij}R^2\, ,               \nn
\eea
one can rederive the forms of $K_{ij}$ found in \cite{1,5}:
\bea
K_{ij}&=&-\frac{1}{2}\nabla_j\nabla_i R+2\nabla^2 L_{ij}
  -\frac{3}{2}\hR_{ij}R-\eta_{ij}K+4R_{imjn}\hR^{mn}           \nn\\
 &=&-\frac{1}{2}\nabla_j\nabla_i R+2\nabla^2 L_{ij}
   +\frac{9}{2}\hR_{ij}R-8\hR_{in}\hR^n{_j}
   +\eta_{ij}\left(3\hR_{mn}\hR^{mn}-\frac{13}{8}R^2\right)\, .\nn
\eea

\section{Reduced phase space formalism}
\setcounter{equation}{0}

Starting from the basic Poisson brackets (PB)
$\{b^i{_\m},\pi_j{^\n}\}=\d^i_j\d^\n_\m\d(\mbox{\boldmath{$x-x'$}})$
etc., one finds that the nontrivial piece of the PB algebra for the
primary constraints $X_A=(\phi_i{}^\a,\Phi_i{}^\a,p_i{}^\a,P_i{}^\a)$
has the form:
\bea
&&\{\phi_i{}^\a,\Phi_j{}^\b\}=-2a\s\ve^{0\a\b}\eta_{ij}\d\, ,
  \qquad\{\phi_i{^\a},p_j^{\b}\}=-\ve^{0\a\b}\eta_{ij}\d\,,\nn\\
&&\{\Phi_i{}^\a,P_j{}^\b\}
  = -\frac{a}{m^2}\ve^{0\a\b}\eta_{ij}\d\, .               \nn
\eea
The consistency conditions of $X_A$ determine the corresponding
multipliers, which are conveniently written in the form:
\bea
&&u{^i}{_\a}=\dot b^i{_\a}
  =-\ve^{ijk}\om_{j0}b_{k\a}+\nab_\a b^i{_0}\, ,          \nn\\
&&v{^i}{_\a}=\dot\om^i{_\a}=\nab_\a\om^i{_0}+
  \frac{1}{2}b\ve_{0\a\b}(\f^{i\b}-\f h^{i\b})\, ,        \nn\\
&&w{^i}{_\a}=\dot\l^i{_\a}=-\ve^{ijk}\om_{j0}\l_{k\a}
  +\nab_\a\l^i{_0} +2\L\ve^{ijk}b_{j0}b_{k\a}
  -a\s b\ve_{0\a\b}(\f^{i\b}-\f h^{i\b})
  +\frac{a}{m^2}b\ve_{0\a\b}\cT_i{^\b}\, ,                  \nn\\
&&z^i{_\a}=\dot\f^i{_\a}=-\ve^{ijk}\om_{j0}\f_{k\a}+\nab_{\a}\f^i{_0}
  -\frac{m^2}{a}\ve^{ijk}(b_{j0}\l_{k\a}-b_{j\a}\l_{k0})\,.\lab{A.1}
\eea

Now, we go over to the reduced phase space $R_1$, defined by
eliminating the momentum variables from the second class constraints
$X_A$. Consider the $24\times 24$ matrix $\D$ with matrix elements
$\D_{AB}=\{X_A,X_B\}$:
\be
\D(\mb{x},\mb{y})=
     \left( \ba{cccc}
        0&            -2a\s &  -1 & 0   \\
      -2a\s &             0 &   0 &-\dis\frac{a}{m^2} \\
      -1 &                0 &   0 & 0   \\
       0 &-\dis\frac{a}{m^2}& 0 & 0
               \ea
     \right)\otimes\ve^{0\a\b}\eta_{ij}\d(\mb{x}-\mb{y})\, .\nn
\ee
The matrix $\D$ is regular, and its inverse has the form
\be
\D^{-1}(\mb{y},\mb{z})=
     \left( \ba{cccc}
            0 &0              &1       & 0      \\
            0 &0              &0       &\dis\frac{m^2}a \\
            1 &0              &0       &-2m^2\s \\
            0 &\dis\frac{m^2}{a}&-2m^2\s &0
            \ea
     \right)\otimes\ve_{0\b\g}\eta^{jk}\d (\mb{y}-\mb{z})\,.\nn
\ee
Thus, the constraints $X_A$ are second class, and $\D^{-1}$ defines
the DBs in $R_1$:
$$
\{\phi,\psi\}^*_1=\{\phi,\psi\}
                  -\{\phi,X_A\}(\D^{-1})^{AB}\{X_B,\psi\}\, .
$$
Explicit form of the nontrivial DBs is displayed in \eq{3.5}.

\section{Dirac brackets}
\setcounter{equation}{0}

In this appendix, we display the set of DBs, needed in the main text.

We start with the DBs of the secondary constraints:
\bsubeq
\bea
&&\{\hcH_i,\hcH_j\}^*_1=4a\bL\ve_{ijk}\cT^k\d              \nn\\
&&\qquad\qquad\quad +\frac{a}{m^2}\left[\frac{\d}{\d b^j{_0}}
   \left(\nab_\a(b\cT_i{^\a})-\frac{b}2\ve_{imn}(\f^{m\a}
   -\f h^{m\a})\l^n{_\a}\right)-(i\leftrightarrow j)\right]\nn\\
&&\{\hcH_i,\cK_j\}^*_1=-\ve_{ijk}\hcH^k\d\, ,              \nn\\
&&\{\hcH_i,\hcR_j\}^*_1= \frac{a}{m^2}
  \left[\frac{\d}{\d\f^j{_0}}\nab_\a(b\cT_i{^\a})
        -\frac{\d}{\d\f^j{_\a}}\nab_\a(b\cT_i{^0})
        -\frac{b}2\ve_{imn}(\d^m{_j}g^{0\a}
        -h_j{^0}h^{m\a})\l^n{_\a}\d \right]                \nn\\
&&\{\hcH_i,\cT_j\}^*_1= -\frac{m^2}{a}\cR^k\d
  +\frac{1}{2}b\ve_{jmn}\left[h_i{^0}\f^{mn}-\f^m{_i}(h^{n0}
  -b^n{_0}g^{00})-\f^0{_i}h^{m0}b^n{_0}\right]\d\,,        \nn\\
&&\{\cK_i,\cK_j\}^*_1 =-\ve_{ijk} \cK^k\d\, ,\qquad
  \{\cK_i,\hcR_j\}_1^*=-\ve_{ijk}\hcR^k\d\, ,              \nn\\
&&\{\cK_i,\cT_j\}^*_1=-\ve_{ijk}\cT^k\d\, ,                \nn\\
&&\{\hcR_i,\hcR_j\}_1^*
  =\left[-\pd_\a(bh_i{^\a}h_j{^0})
   +b\ve_{imn}\om^m{_\a}(g^{0\a}\d^n{_j}-h^{n\a}h_j{^0})
   \right]\d-(i\leftrightarrow j)\, ,                      \nn\\
&&\{\hcR_i,\cT_j\}_1^*=\frac{1}{2}b\ve_{jmk}
  (\d^m_i g^{0\b}-h_i{^0}h^{m\b})b^k{_\b}\d\, ,           \nn\\
&&\{\cT_i,\cT_j\}^*_1=0\, .
\eea
The DBs between the secondary first class constraints are given by:
\bea
&&\{\bcH_i,\bcH_j\}=-\ve_{ijk}(\f^{kn}-\f\eta^{kn})\bcK_n\d\,,\nn\\
&&\{\bcH_i,\bcK_j\}=-\ve_{ijk}\bcH^k\d\, ,                 \nn\\
&&\{\bcK_i,\bcK_j\}=-\ve_{ijk}\bcK^k\d\, .
\eea
\esubeq
Finally, we display the most important DBs involving the tertiary
constraints:
\bea
&&\{\cT_i,\ve^{0\a\b}\f_{\a\b}\}*_1
  =-\frac{m^2}{a}\ve^{0\a\b}\ve_{ijk}b^j{_\a}b^k{_\b}\d
  \equiv -\frac{m^2}{a}2b h_i{^0}\d\, ,                    \nn\\
&&\{\cT_i,\ve^{0\a\b}\l_{\a\b}\}*_1=\cT_i\d\approx 0\, ,   \nn\\
&&\{\hcR_i,\ve^{0\a\b}\f_{\a\b}\}^*_1=\cT_i\d\approx 0\, , \nn\\
&&\{\hcR_i,\ve^{0\a\b}\l_{\a\b}\}^*_1
  =\frac{ab}{2m^2}\left[\f_i{^0}-\f h_i{^0}
   +g^{00}(\f_{i0}+2m^2\s b_{i0})
   -h_i{^0}(\f^0{_0}-2m^2\s)\right]\d\,,                   \nn\\
&&\{\ve^{0\a\b}\l_{\a\b},\ve^{0\g\d}\f_{\g\d}\}^*_1
  =-\ve^{0\a\b}\f_{\a\b}\d\approx 0\, .
\eea

\section{Two identities}
\setcounter{equation}{0}

In this appendix, we prove that equations \eq{3.10a} and \eq{3.12b} are
identities.

\prg{1.} Equation \eq{3.10a} can be rewritten in the following form:
\be
Z'^\a{_\a}+\f_\b{^\b}U^\a{_\a}+U^0{_\a}\f_0{^\a}=
  b^{-1}b^i{_0}\nab_\a
               \left[b(\f_i{^\a}-\f h_i{^\a})\right]\, .   \lab{D.1}
\ee
Using the relations:
\bea
&&Z^\a{_\a}=h_i{^\a}\nab_\a \f^i{_0}\, ,                   \nn\\
&&Z'{}^\a{_\a}= h_i{^\a}\nab_\a\f^i{_0}-\f^\a{_i}\nab_\a b^i{_0}
   \approx b^i{_0}\nab_\a \f_i{^\a}-\f^i{_0}\nab_\a h_i{^\a}\,,\nn
\eea
the left-hand side of \eq{D.1} can be transformed into
\be
L= b^i{_0}\nab_\a \f_i{^\a}-\f^i{_0}\nab_\a h_i{^\a}
  -\f_\b{^\b}b^i{_0}\nab_\a h_i{^\a}
  +\f_0{^\a}h_i{^0}\nab_\a b^i{_0}\, .                     \nn
\ee
Let us now rewrite the right-hand side of \eq{D.1} as
$$
R=b^i{_0}\nab_\a\f_i{^\a}-\f b^i{_0}\nab_\a h_i{^\a}
  +\f_0{^\a} b^{-1}\nab_\a b\, .
$$
By noting that
\bea
&&b^{-1}\nab_\a b=
  h_i{^0}\nab_\a b^i{_0}+h_i{^\b}\nab_\a b^i{_\b}\, ,      \nn\\
&&\f_0{^\a}h_i{^\b}\nab_\a b^i{_\b}
  \approx -\f_0{^\a} b^i{_\a}\nab_\b h_i{^\b}
  \approx -(\f^i{_0}-\f_0{^0}b^i{_0})\nab_\a h_i{^\a}\, ,  \nn
\eea
we find $R\equiv L$, which implies that equation \eq{3.10a} is
satisfied identically.

\prg{2.} The consistency condition \eq{3.12b} can be rewritten
in the form:
\bsubeq
\be
Z'_{0\b}=Z'_{\b0}\, .                                      \lab{D.2a}
\ee
Since $Z'_{\b 0}$ is determined from \eq{3.10b}, the proof that
\eq{D.2a} is an identity is realized by showing that the substitution
of \eq{D.2a} into \eq{3.10b} yields an identity. By making use of
\eq{D.2a} and $Z'_{\b\a}=Z'_{\a\b}$, equation \eq{3.10b} takes the
following form:
\bea
Z'^0{_\b}+\f_\b{^0}U^\a{_\a}-(\f_\b{^\a}-\f\d^\a_\b)U^0{_\a}
  +b^{-1}b^i{_\b}\nab_\a\left[b(\f_i{^\a}-\f h_i{^\a})\right]=0\,.\lab{D.2b}
\eea
\esubeq
Let us now use the following relations:
\bea
&&Z'^0{_\b}=h^{i0}\nab_\b \f_{i0}-\f^0{_i}\nab_\b b^i{_0}
  -\frac{m^2}{a}\ve_{ijk}\left(h^{i0}b^j{_0}\l^k{_\b}
                          -h^{i0}b^j{_\b}\l^k{_0}\right)\,,\nn\\
&&\f_\b{^0}U^\a{_\a}-(\f_\b{^\a}-\f\d^\a_\b)U^0{_\a}=
  -\f_\b{^0}b^i{_0}\nab_\a h_i{^\a}
  +(\f_\b{^\a}-\f\d^\a_\b)b^i{_0}\nab_\a h_i{^0}\, ,       \nn\\
&&b^{-1}b^i{_\b}\nab_\a\left[b(\f_i{^\a}-\f h_i{^\a})\right]=
  (b^{-1}\nab_\a b)(\f_\b{^\a}-\f \d^\a_\b)+b^i{_\b}\nab_\a\f_i{^\a}
  -\pd_\b\f-\f b^i{_\b}\nab_\a h_i{^\a}\, ,                \nn
\eea
and, in addition to that,
\be
(b^{-1}\nab_\a b)(\f_\b{^\a}-\f \d^\a_\b)=
  -(b^i{_0}\nab_\a h_i{^0})(\f_\b{^\a}-\f\d^\a_\b)-(\nab_\a h_i{^\a})
   \left(\f_\b{^i}-b^i{_0}\f_\b{^0}-\f b^i{_\b}\right)\,.  \nn
\ee
Then, the left-hand side of \eq{D.2b} takes the following form:
\bea
&&h^{i\m}\nab_\b \f_{i\m}-\pd_\b\f-\f^{i\m}\nab_\b b_{i\m}
  +h_i{^\a}(\nab_\a \f^i{_\b}-\nab_\b \f^i{_\a})
  -\frac{m^2}{a}\ve_{ijk}\left(h^{i0}b^j{_0}\l^k{_\b}
                              -h^{i0}b^j{_\b}\l^k{_0}\right)\approx\nn\\
&&-\f_{i\m}\nab_\b h^{i\m}-\f^{i\m}\nab_\b b_{i\m}
  -\frac{m^2}a\ve_{ijk}\left(h^{i\m}b^j{_\m}\l^k{_\b}
  -h^{i\m}b^j{_\b}\l^k{_\m}\right)\approx                  \nn\\
&&(\f^{\m i}-\f^{i\m})\nab_\b b_{i\m}\approx 0\,.          \nn
\eea
Hence, relation \eq{D.2b} is satisfied identically, which completes our
proof.

\section{Second class constraints}
\setcounter{equation}{0}

In this appendix, we show that the set of 20 constraints in the second
column of Table 1 are second class. Instead of calculating the
determinant of the $20\times 20$ matrix of the related DBs, the proof
is derived iteratively.

\prg{Step 1.} We begin by considering the subset of constraints
$Y_A:=(\th_{0\b},\ups,P^{\a 0}, P_0{^0})$. The $6\times 6$ matrix
$\D_1$ with matrix elements $\{Y_A,Y_B\}^*_1$ has the form:
\bea
&&\D_1=\left(\ba{cc}
             0_{3\times 3} & A_{3\times 3} \\[5pt]
            -A^T_{3\times 3} & 0_{3\times 3}
             \ea\right)\, ,                                \nn\\[5pt]
&&A:=\left(\ba{cc}
        \{\th_{0\a},P^{\b0}\}_1^* &\{\th_{0\a},P_0{^0}\}^*_1  \\
          \{\ups,P^{\b0}\}_1^*&\{\ups,P_0{^0}\}_1^*
       \ea\right)\, .                                      \nn
\eea
The explicit form of $A$ reads:
\be
A=\left(\ba{cc}
         -\d_\a^\b  & -g_{0\a}  \\
          \Om^{0\b} & \Om_0{^0}
         \ea\right)\d\, .                                  \nn
\ee
Using the formulas
\bea
&&\det A= g_{00}\Om^{00}\, ,                               \nn\\
&&\det\D_1= \det(AA^T)=\left(\det A\right)^2\, ,
\eea
we see that $\D_1$ is regular provided the condition \eq{3.16b} is
satisfied.

\prg{Step 2.} Next, we focus our attention on the subset
$Z_A:=(\psi_{0\b},\chi,p^{\a 0}, p_0{^0})$. The corresponding $6\times
6$ matrix $\D_2$ reads:
$$
\D_2=\left(\ba{cc}
           B_{3\times 3} & C_{3\times 3} \\[5pt]
          -C^T_{3\times 3} & 0_{3\times 3}
          \ea\right)\, ,
$$
where
\bea
&&B:=\left(\ba{cc}
    -2\ve_{0\a\b}\l_{00}&-2\ve_{0\a\g}\l^\g{_0} \\
        2\ve_{0\b\g}\l^\g{_0} & 0
           \ea\right)\d\, ,                                \nn\\[5pt]
&&C:=\left(\ba{cc}
           -\d_\a^\b & -g_{0\a} \\
            g^{0\b}& 1
           \ea\right)\d\, .                                \nn
\eea
The matrix $\D_2$ is regular:
\bea
&&\det(C)=g^{00}g_{00}\, ,                                 \nn\\
&&\det\D_2=(\det C)^2\, .
\eea

\prg{Step 3.} Finally, we consider the remaining subset
$W_A=(\cT_i,\ve^{0\a\b}\l_{\a\b},\hcR'_i,\ve^{0\a\b}\f_{\a\b})$; these
constraints do not contain the variables $\f_{i0},\l_{i0}$. The
$8\times 8$ matrix $\{W_A,W_B\}^*_1$ takes the form
$$
\D_3=\left(\ba{cc}
           0_{4\times 4} & M_{4\times 4} \\[9pt]
          -M^T_{4\times 4} & N_{4\times 4}
           \ea\right)\, ,
$$
where
$$
M=\left(\ba{cc}
        D_{3\times 3} & E_{3\times 1} \\[9pt]
       -H^T_{1\times 3} & 0_{1\times 1}
           \ea\right)\, ,\qquad
N=\left(\ba{cc}
        F_{3\times 3} & 0_{3\times 1} \\[9pt]
        0_{1\times 3} & 0_{1\times 1}
           \ea\right)\, ,
$$
and the matrices $D,E,F$ and $H$ are given by
\bea
D_{ij}&:=&\{\cT_i,\hcR'_j\}^*_1=-b\left[\ve_{ijn}\left(
  \frac{1}{2}h^{n0}-g^{00}b^n{_0}\right)
  -h_j{^0}\ve_{imn}b^m{_0}h^{n0}\right]\d\, ,              \nn\\
E_i&:=&\{\cT_i,\ve^{0\a\b}\f_{\a\b}\}^*_1
      =-\frac{m^2}a 2bh_i{^0}\d\, ,                        \nn\\
H_i&:=&\{\hcR'_i,\ve^{0\a\b}\l_{\a\b}\}^*_1=2ab\Om_i{^0}\d\,,\nn\\
F_{ij}&:=&\{\hcR'_i,\hcR'_j\}_1^*\, .                      \nn
\eea
The calculation of $\det\D_3$ yields
\bea
&&\det M=\frac{1}{2}\ve^{ijk}\ve^{mnl}D_{im}D_{jn}E_kH_l
        =-m^2b^4g^{00}\Om^{00}\, ,                         \nn\\
&&\det\D_3=(\det M)^2\, .                                  \nn
\eea
Thus, $\det\D_3\ne 0$ provided  $\Om^{00}\ne 0$.


\end{document}